\documentclass{ws-procs9x6}

\begin{document}

\title{Determining the $\Theta^+$ quantum numbers through 
a Kaon induced reaction}

\author{T.~HYODO and A.~HOSAKA}

\address{Research Center for Nuclear Physics (RCNP), \\
Ibaraki, Osaka 567-0047, Japan \\ 
E-mail: hyodo@rcnp.osaka-u.ac.jp}

\author{E.~OSET and M.~J.~VICENTE VACAS}

\address{Departmento de F\'isica Te\'orica and IFIC, \\
Centro Mixto Universidad de Valencia-CSIC, \\
Institutos de Investigaci\'on de Paterna, Aptd. 22085, 46071
Valencia, Spain}  

\maketitle

\abstracts{
    We study the $K^+p\to \pi^+KN$ reaction with kinematical
    condition suited to the production of the $\Theta^+$ resonance.
    It is shown that in this reaction
    with the polarization experiment,
    a combined consideration of the strength
    at the peak and the angular dependence of cross section
    can help determine the $\Theta^+$ quantum numbers.
}


In 2003, the LEPS collaboration
reported a signal for a narrow $S=+1$ resonance
around 1540 MeV in the experiment held at SPring-8.\cite{Nakano:2003qx}
It is important  for further theoretical studies
to determine experimentally the quantum numbers of the
$\Theta^+$ resonance such as spin and parity.
Here we study the $K^+p\to \pi^+KN$ reaction with
different $\Theta^+$ quantum
numbers, and see qualitative differences in observables 
depending on the quantum
numbers.\cite{Hyodo:2003th,Oset:2003wu,Oset:2004up}
The $K^+p\to \pi^+KN$  reaction was also studied theoretically in
Refs.~\refcite{Liu:2003rh,Oh:2003gj}, but in the present study,
we take the background contribution into account, which should
be important for the information on the signal/background ratio
and the possible interference between them.

\begin{figure}[b] 
    \epsfxsize=11.5cm   
    \centerline{
    \epsfbox{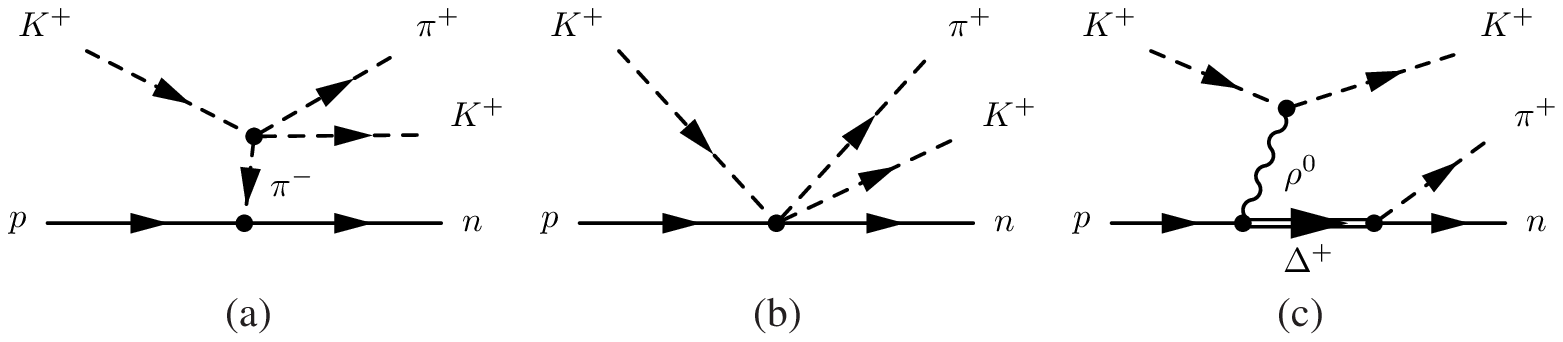}
    }
    \epsfxsize=8.5cm   
    \centerline{
    \epsfbox{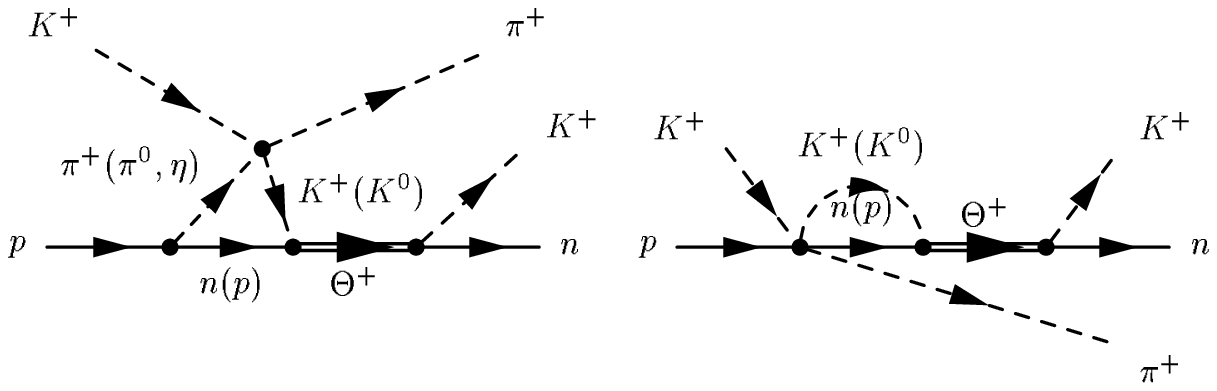}
    }
    \vspace{-0.5cm}
    \caption{Upper panel : Feynman diagrams of the reaction $K^+p\to \pi^+K^+n$
    in a model.
    Lower panel : the $\Theta^+$ resonance contribution.
    } 
\label{fig:1} 
\end{figure}

A successful model for the reaction was
considered in Ref.~\refcite{Oset:1996ns},
consisting of the mechanisms depicted in
terms of Feynman diagrams in the upper panel of Fig.~\ref{fig:1}.
The terms (a) (meson pole) and (b) (contact term) are derived
from the effective chiral Lagrangians.
Since the term (c) is proportional to the momentum of the final pion,
it is negligible when the momentum is small.
In the following, we calculate
in the limit of the pion momentum set to zero.
If there is a resonant state for $K^+n$ then this 
will be seen in the final state interaction of this system.
These processes are expressed as in the lower panel of Fig.~\ref{fig:1},
and they contribute to the present reaction 
in addition to the diagrams (a) and (b).
For an $s$-wave $K^+n$ resonance we have $J^P=1/2^-$,
and for a $p$-wave, $J^P=1/2^+, 3/2^+$.
We write the couplings of the resonance to $K^+n$ as $g_{K^+n}$,
$\bar{g}_{K^+n}$ and $\tilde{g}_{K^+n}$ 
for $s$-wave and $p$-wave with $J^P=1/2^+,3/2^+$ respectively,
and relate them
to the $\Theta^+$ width $\Gamma=20$ MeV via
$g_{K^+n}^2=\frac{\pi M_R\Gamma}{Mq}$,
$\bar{g}_{K^+n}^2
=\frac{\pi M_R\Gamma}{Mq^3}$ and
$\tilde{g}_{K^+n}^2=\frac{3\pi M_R\Gamma}{Mq^3}$.

A straightforward evaluation of the diagrams $(a)$ and $(b)$ 
leads to the $K^+n\to \pi^+KN$
amplitudes 
\begin{equation}
    -it_i
    =a_i\vec{\sigma}
    \cdot\vec{k}_{in}
    +b_i\vec{\sigma}
    \cdot\vec{q}^{\prime} \ ,
    \label{eq:t1amp}
\end{equation}
where
the explicit form of coefficients $a_i,b_i$ are given in 
Ref.~\refcite{Hyodo:2003th},
$i=1,2$ stands for the final state $K^+n, K^0p$ respectively,
and $k_{in}$ and $q^{\prime}$ are the initial and final $K^+$ momenta.
With some coefficients $f_i$, $d_i$,\ldots,
the resonance terms $\tilde{t}_i$ take on the form
\begin{equation}
    -i\tilde{t}^{(s)}_i
    =c_i
    \vec{\sigma}\cdot\vec{k}_{in} \ , \quad
    -i\tilde{t}^{(p,1/2)}_i
    =d_i
    \vec{\sigma}\cdot\vec{q}^{\prime}  \ , \quad
    -i\tilde{t}^{(p,3/2)}_i
    =f_i\vec{\sigma}\cdot\vec{k}_{in}
    -g_i\vec{\sigma}\cdot\vec{q}^{\prime} \ . 
    \label{eq:tilde}
\end{equation}
where the subscript $i$ accounts for the intermediate state
and the upper superscripts denote the partial wave and
the spin of the $\Theta^+$.
Finally the total amplitude for $K^+n$ final state is given by
\begin{equation}
    -i\tilde{t}=-it_1-i\tilde{t}_1-i\tilde{t}_2
    \label{eq:total}
\end{equation}

\begin{figure}[tb] 
    \epsfxsize=7cm   
    \centerline{
    \epsfbox{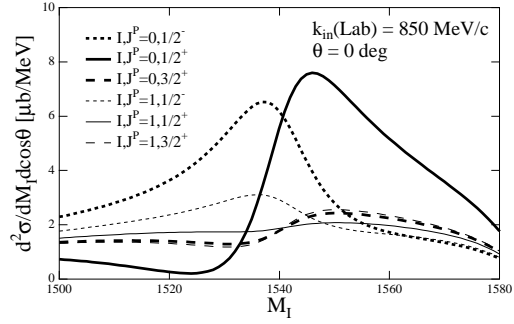}
    }
    \caption{The double differential cross sections 
    at $\theta=0$ deg (forward direction)
    for $I=0,1$ and $J^P=1/2^-,1/2^+,3/2^+$.
    } 
\label{fig:2} 
\end{figure}

We calculate the cross sections with
the initial three momentum of $K^+$ in the Laboratory frame
$k_{in}(Lab)=850$ MeV$/c$ ($\sqrt{s}=1722$ MeV).
At this energy, the final $\pi^+$ momentum
is small enough with respect to $|\vec{k}_{in}|$.
In Fig.~\ref{fig:2},
we show the invariant mass distribution
$d^2\sigma/dM_Id\cos\theta$ in the $K^+$
forward direction ($\theta=0)$.
A resonance signal is always observed,
independently of the quantum numbers of $\Theta^+$.
The signals for the resonance are quite clear for the case
of $I,J^P=0,1/2^+$  and  $I,J^P=0,1/2^-$, while in the other
cases the signals are weaker and the background is more important.

\begin{figure}[tb] 
    \epsfxsize=7cm   
    \centerline{
    \epsfbox{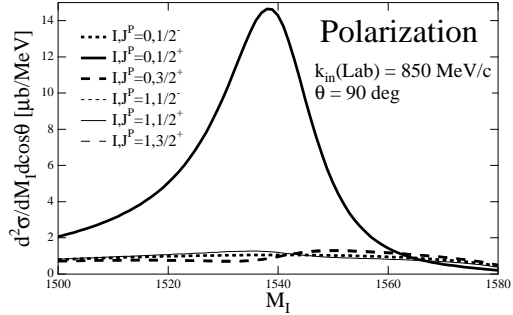}
    }
    \caption{The double differential cross sections 
    at $\theta=90$ deg
    for $I=0,1$ and $J^P=1/2^-,1/2^+,3/2^+$,
    with polarized initial photon.
    } 
\label{fig:3} 
\end{figure}

Next we consider polarized amplitudes.
It is seen in Eqs.~(\ref{eq:tilde}) that if the $\Theta^+$ 
has the negative parity,
the amplitude is proportional to $\vec{\sigma}\cdot \vec{k}_{in}$ 
while if it has positive parity with spin $1/2$,
it is proportional to
$\vec{\sigma}\cdot \vec{q}^{\prime}$.
We try to use this property in order to distinguish the two cases.
Let us consider that the initial proton polarization is $1/2$
in the direction $z$ $(\vec{k}_{in})$
and the final neutron polarization is $-1/2$  (the experiment
can be equally done with $K^0p$ in the final state, which makes the
nucleon detection easier).
In this spin flip amplitude $\langle -1/2 |t|+1/2\rangle$ the 
$\vec{\sigma}\cdot \vec{k}_{in}$ term vanishes, and
therefore  the resonance signal disappears for the $s$-wave
case, while the $\vec{\sigma}\cdot \vec{q}^{\prime}$ operator of the
$p$-wave case would have a finite matrix element 
proportional to $q^{\prime}\sin \theta$.
This means, away from the forward direction of the final kaon, the
appearance of a resonance peak in the mass distribution would indicate a
$p$-wave coupling and hence a positive parity resonance.
In  Fig.~\ref{fig:3},
we show the results for the polarized cross section
measured at 90 degrees as a function of the invariant mass.
The absence of the resonance term in $s$-wave results is clearly seen.
The only sizeable resonance peak comes from the $I,J^P=0,1/2^+$ case,
and the other cross sections for spin $3/2$ are quite reduced.
A clear experimental signal of the resonance in this observable
would indicate the quantum numbers as  $I,J^P=0,1/2^+$.

In summary, we have studied the $K^+p \to \pi^+ KN$ reaction
and shown a method to determine the quantum numbers of $\Theta^+$
in the experiment, by using the polarized cross sections.
For future perspective, 
the calculation for a finite momentum of final pion 
is desired. A combined study of $(K^+,\pi^+)$ and $(\pi^-,K^-)$
reactions based on two-meson coupling\cite{Hosaka:2004mv}
is also in progress.\cite{reaction}

\section*{Acknowledgments}
This work is supported by the Japan-Europe (Spain) Research
Cooperation Program of Japan Society for the Promotion of Science
(JSPS) and Spanish Council for Scientific Research (CSIC), which
enabled T.~H. and A.~H. to visit
IFIC, Valencia and E.~O. and M.~J.~V.~V. visit RCNP, Osaka.
This work is also supported in part  by DGICYT
projects BFM2000-1326,
and the EU network EURIDICE contract
HPRN-CT-2002-00311.

%
%
%
%

%
%

\end{document}